\documentclass[multphys,vecphys]{svmult}

\usepackage{makeidx}         
\usepackage{graphicx}        
\usepackage{multicol}        
\usepackage[bottom]{footmisc}

\makeindex             

\begin{document}

\title*{A Heating Model for the Millennium Gas Run}
\author{Lorena Gazzola\and
Frazer R. Pearce}
\institute{School of Physics and Astronomy, University of Nottingham, NG7 2RD, UK
\texttt{ppxlg@nottingham.ac.uk}
}
%
%
\maketitle
 
\begin{abstract}
The comparison between observations of galaxy clusters thermo-dyna-\\mical properties and theoretical predictions suggests that non-gravitational heating needs to be added into the models. We implement an internally self-consistent heating scheme into {\small GADGET-2} for the third (and fourth) run of the Millennium gas project (Pearce et al. in preparation), a set of four hydrodynamical cosmological simulations with $N=2\times(5\times10^8)$ particles and with the same volume ($L=500  h^{-1} \rm{Mpc}$) and structures as the the N-body Millennium Simulation (Springel et al. 2005). Our aim is to reproduce the observed thermo-dynamical properties of galaxy clusters.
\end{abstract}

\section{Model}
\label{sec:2}
The large dynamical range that characterises cosmological simulations and the fact that the physics of heating mechanisms like AGN feedback, galactic winds and conduction has typical scales much smaller that those describing galaxy clusters (few $\rm{kpc}$ versus $\rm{Mpc}$), make their implementation challenging. In addition the Millennium gas runs do not have a high enough resolution to properly model these phenomena ($m_{gas} =3.12 \times 10^9h^{-1}M_{\odot}$, softening $25 h^{-1} \rm{kpc}$) and therefore we need to adopt a relatively simple heating scheme. Our model seeks to improve on the simple preheating scheme that was implemented in the second run of the Millennium gas project.\\
We implement a self-regulated star formation plus feedback procedure by selecting gas particles to convert into stars by imposing a temperature plus over-density threshold (in units of critical density). We choose an over-density rather that a physical density threshold. We then inject an energy $E_{inj}$ to the neighbours of the new star, weighted in distance by the standard SPH smoothing kernel.\\
We choose a temperature threshold of $10^{5}K$ in both models and density threshold, $\rho_{thr}$, to be:
\begin{itemize}
\item  $\rho_{thr} = 200$: mimics the energy deposited by supernovae in galaxy cluster outskirts, low mass objects and at early times
\item  $\rho_{thr} = 2500$: mimics the effect of star formation in the centre of galaxy clusters
\end{itemize}
We then tune $E_{inj}$ in order to reproduce the observed luminosity-temperature relation of galaxy clusters at $z=0$.

\begin{figure}
\centering
\includegraphics[height=5.2cm]{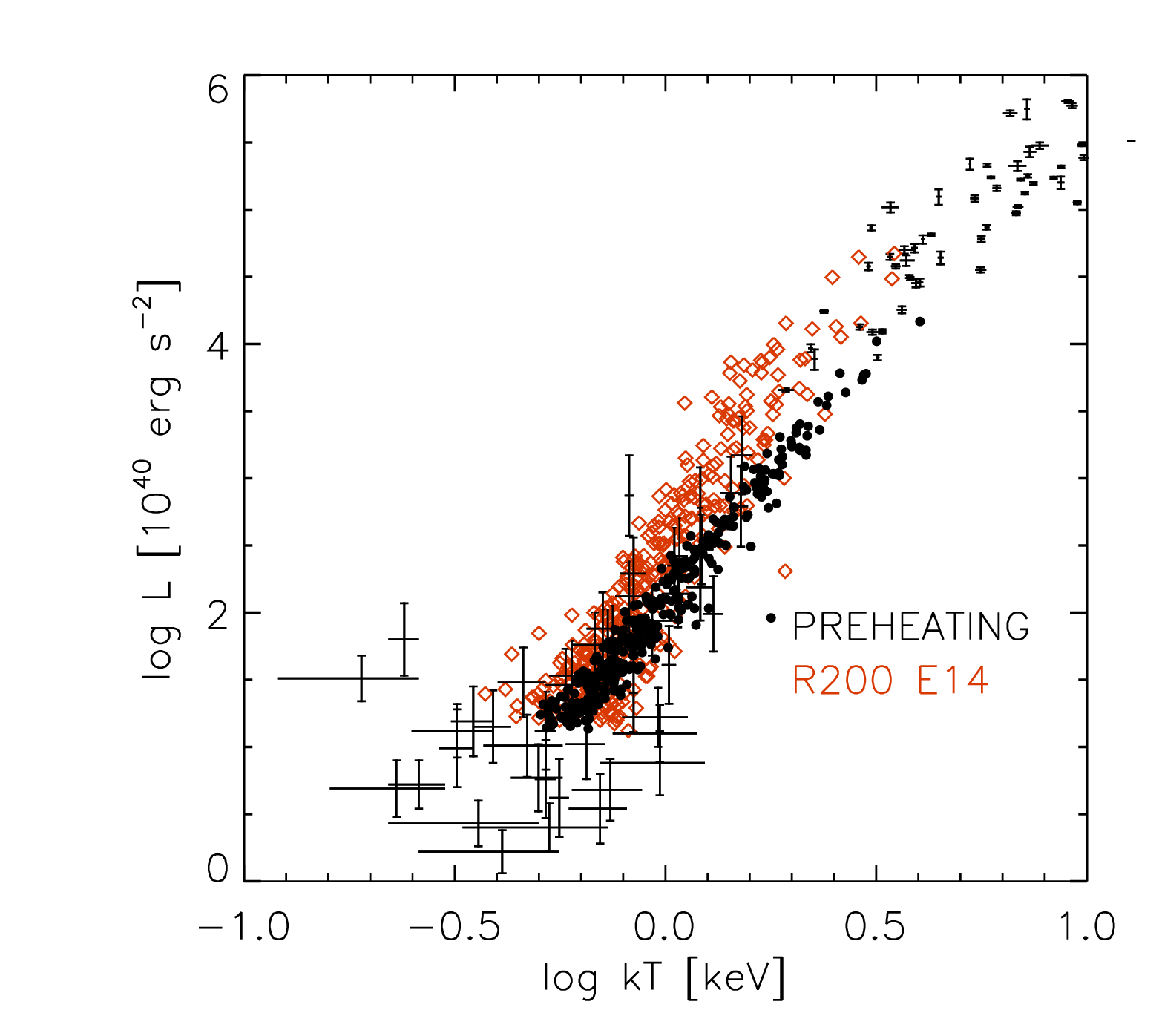}
\includegraphics[height=5.2cm]{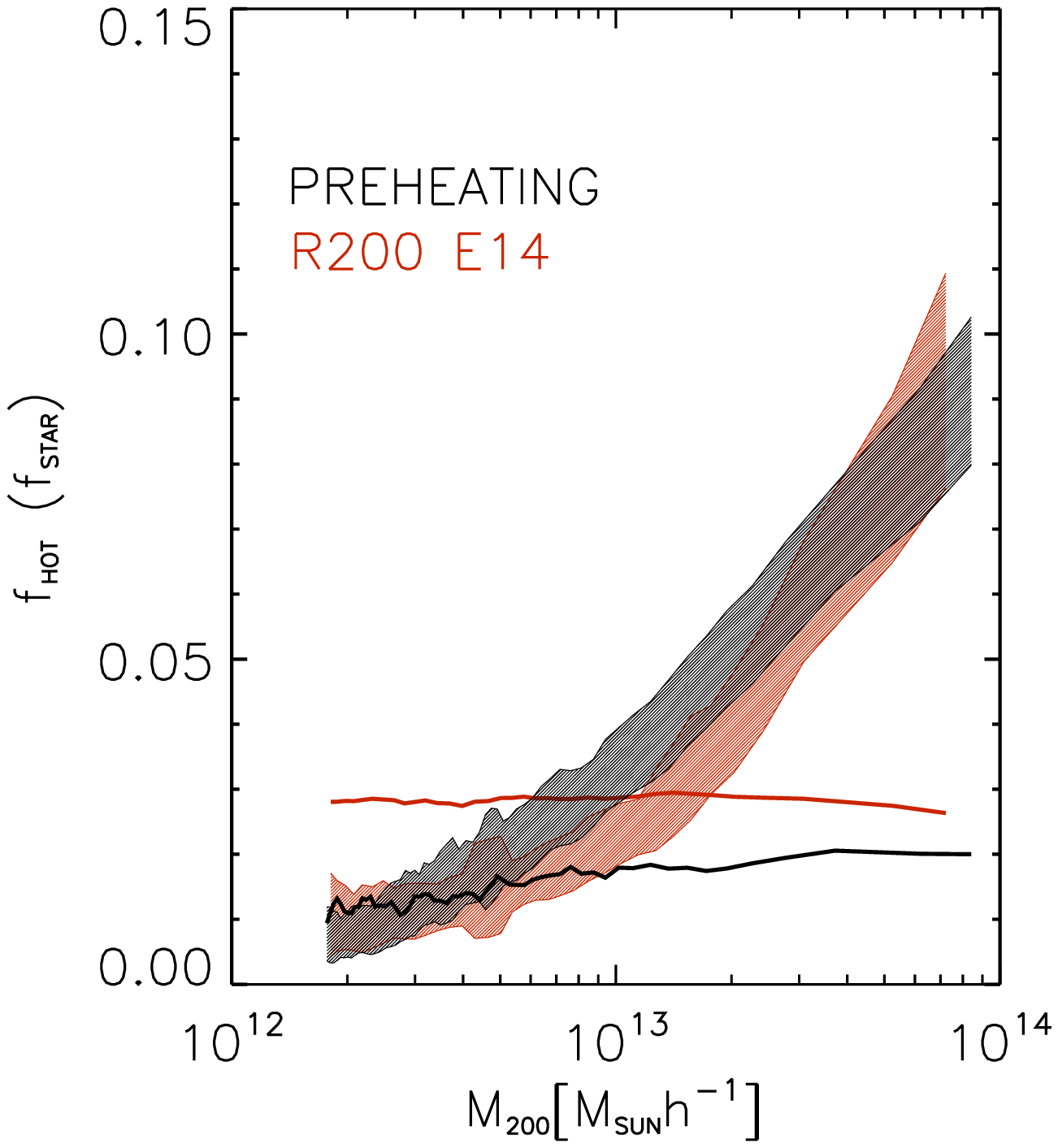}
\caption{Left panel: bolometric luminosity versus mass weighted temperature for groups and clusters in a test box of size $L=125 h^{-1} \rm{Mpc}$. Black dots are for the preheating run, red diamonds for $\rho_{thr}=200$. Observations from Ponman et al. 1996; Helsdon et al. 2000 and  Novicki et al. 2002. Right panel: hot gas (shaded area) and stellar fraction (lines) for the preheating run (black lines) and $\rho_{thr}=200$ (red lines).
\label{fig1}}
\end{figure}
\section{Results}
\begin{figure}
\centering
\includegraphics[height=3.6cm]{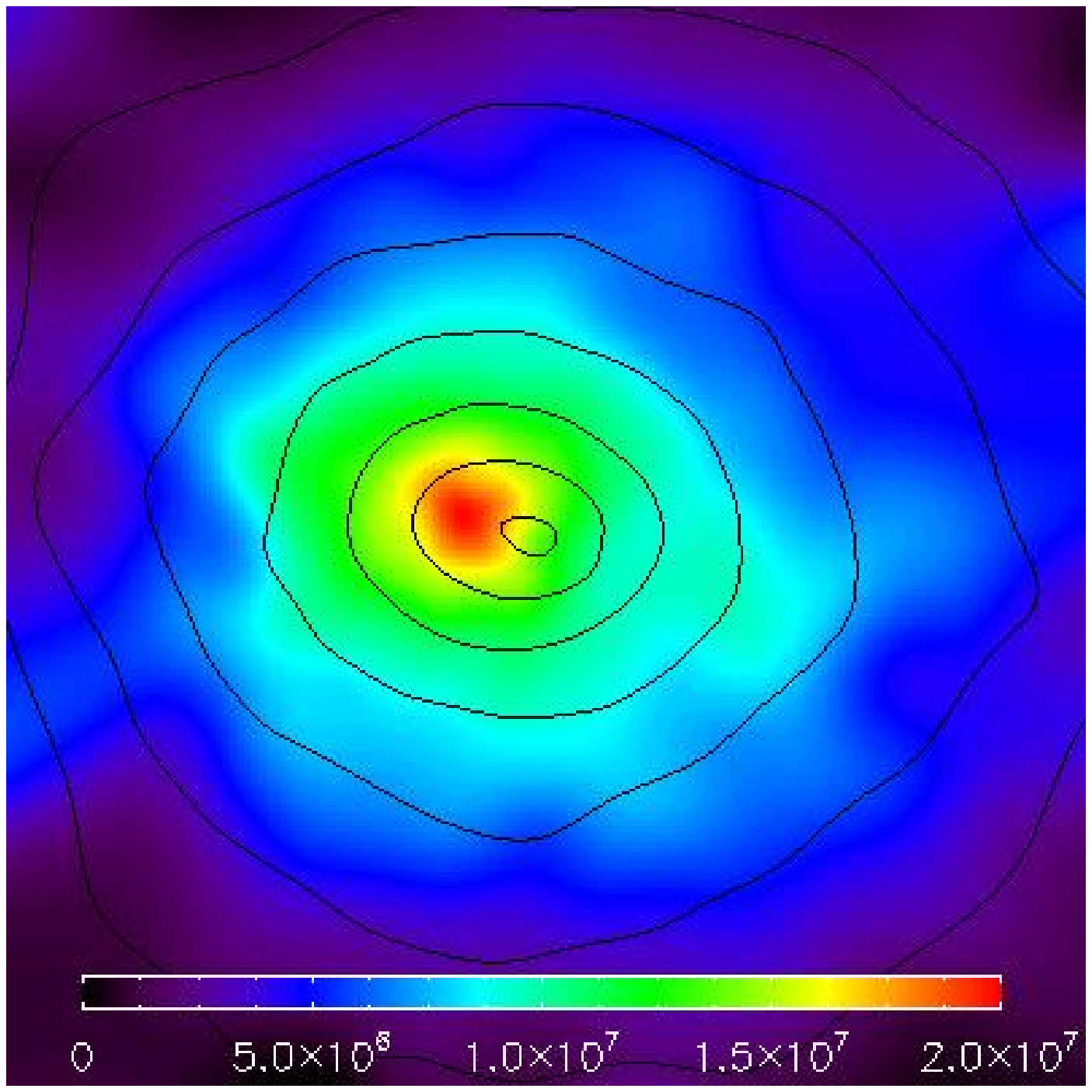}
\includegraphics[height=3.6cm]{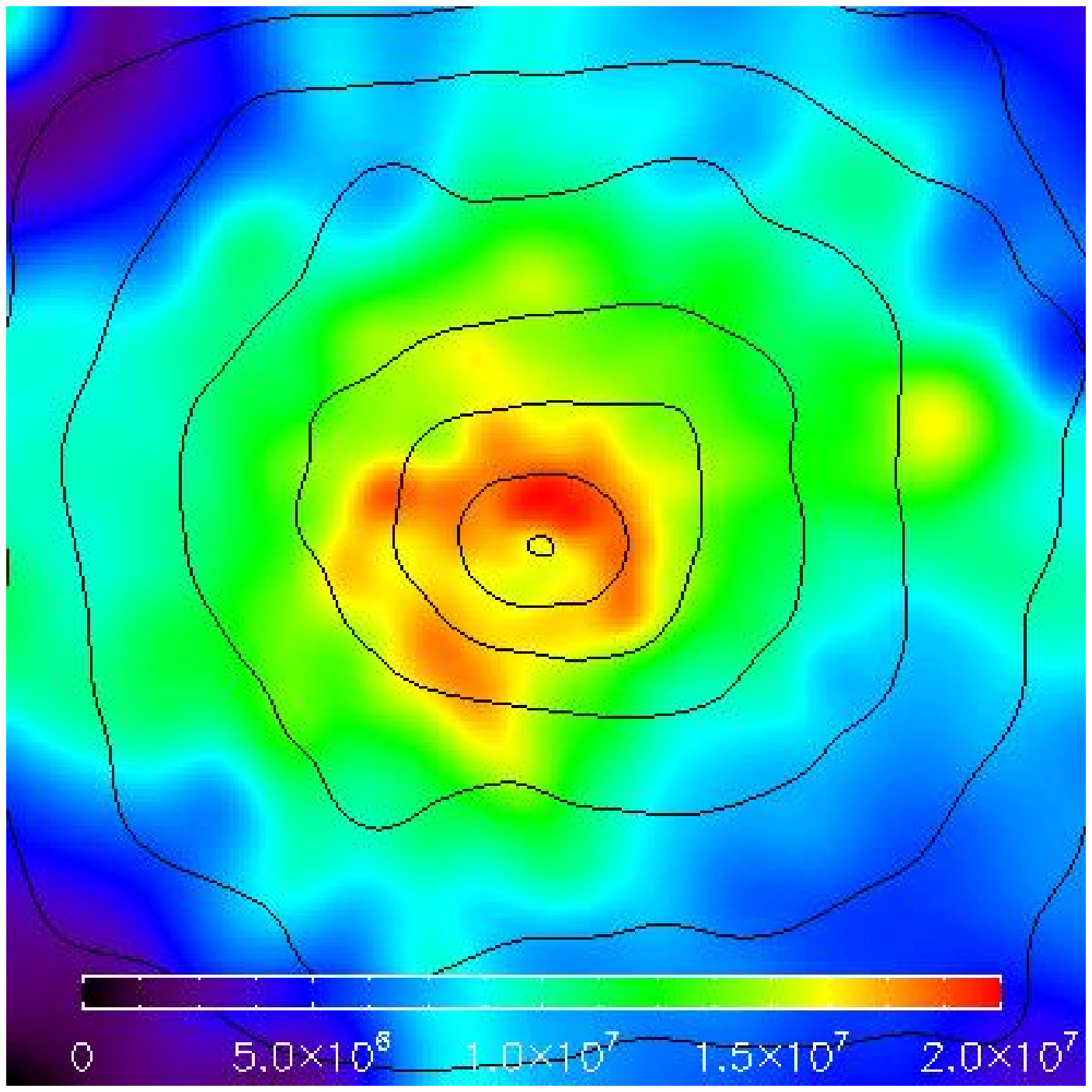}
\includegraphics[height=3.6cm]{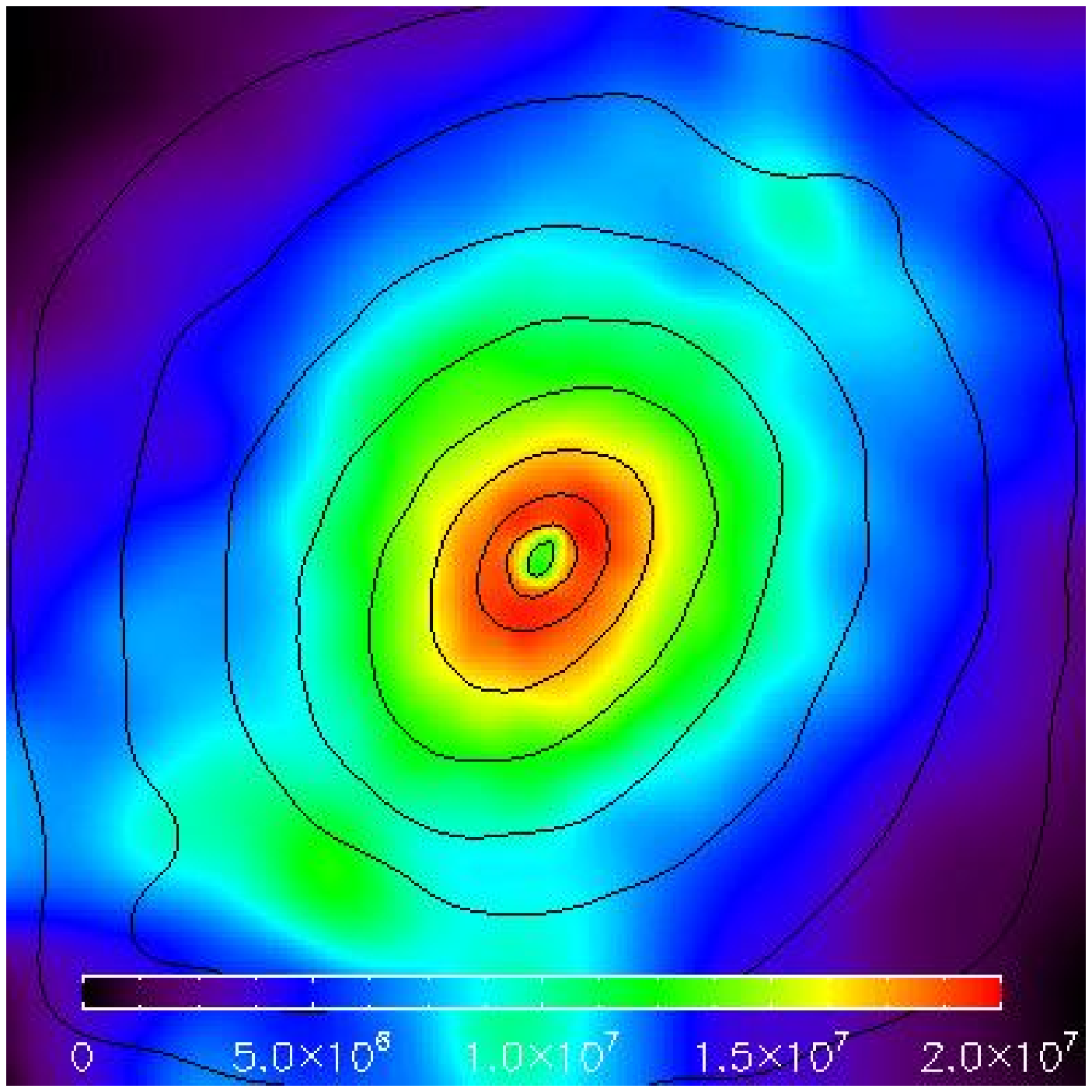}

\caption{Mass weighted temperature maps for three clusters. The panel size is four times the virial radius. Superimposed surface brightness contours.
\label{fig2}}       
\end{figure}

The amount of energy that we inject, $E_{inj}$, is our free parameter and it will be different for the two runs, which vary the density of gas into which the energy is injected. From a test run with $L=250 h^{-1} \rm{Mpc}$ we found that choosing $E_{inj}=14keV$ for $\rho_{thr} = 200$ we manage to reproduce the observed $L-T$ relation for group size objects as well as clusters in the range of $1-6keV$. Comparing this run with the preheating model (similar to the model proposed by Kay et al. 2004), we find that the two runs have a slightly different normalisation but still in agreement with observations (Figure \ref{fig1}). We also notice that the $L-T$ relation has too little scatter in the preheating run, compared to the observations, while the current model appears far more reasonable. This difference is at least partially due to the difference in the temperature structure: in the preheating run the gas is brought to such a high temperature at early times that no cool gas is found at redshift zero and the haloes are characterised by a very smooth gas distribution. On the contrary cool cores can form in our current model and the clusters often show multiple-structure in temperature maps and offsets between the emission peak and the temperature peak (Figure \ref{fig2}).\\

In the second panel of Figure \ref{fig1} we show the hot gas fraction and the star fraction as a function of cluster mass. Tuning the preheating run and $\rho_{thr} = 200$ to match the observed $L-T$ relation at $z=0$, results in a much reduced baryon fraction relative to the non-cooling case, with $f_{bar} \le 0.13$ while the universal fraction is $0.18$ for this cosmology. It is interesting to notice that we obtain similar trends despite the two very different heating mechanisms: one is an external heating (preheating run) and the other one an internal heating. Once the Millennium volume will be available we can use the simulations to test how powerful $f_{hot}(z)$ -hot gas fraction- and $f_{bar}(z)$ -baryonic fraction- are as cosmological probes.\\
The full $500 h^{-1} \rm{Mpc}$ millennium run will also be suitable for time evolution studies. The major drawback of the preheating scheme is that the gas is brought to such a high adiabat at z=4, when the preheating occurs, that no gas can condense any more and star formation is quenched. With this alternative model we get a more realistic star formation history and time evolution. 

\section{Conclusions}
We present an alternative to a simple preheating model by introducing a self-regulated star formation plus feedback scheme. 
The improvements of this model relative to the preheating one are: an increased scatter in the luminosity-temperature relation, the existence of cool core clusters and a more realistic time evolution.
On the other hand the amount of energy required in order to match the observational data with this scheme is quite high, especially if we employ a high density threshold that has been tuned to match the observed $L-T$ at $z=0$.
We are currently exploring higher and lower resolution simulations to constrain the systematic effects introduced by our model.

%
%

%

\begin{thebibliography}{99.}
%
%
%


\bibitem{H00} Helsdon, S. F., \& Ponman, T. J. 2000a, MNRAS, 315, 356
\bibitem{K04} Kay, S. T., Thomas, P. A., Jenkins, A. and Pearce, F. R. 2004b, astro-ph/0411650
A \textbf{61}, 33 (1995)
\bibitem{N02} Novicki, M. C., Sornig, M., \& Henry, J. P. 2002, AJ, 124, 2413 
\bibitem{P96} Ponman, T. J., Bourner, P. D. J., Ebeling, H., \& Bohringer, H. 1996, MNRAS, 283, 690
\bibitem{S05} Springel V., White S. D. M., Jenkins A., et al., 2005b, Nature, 435, 629
\bibitem{P06} Pearce, F. R, Gazzola, L., Kay, S., Thomas, P. A. 2006, in preparation


\end{thebibliography}
%



\printindex
\end{document}